\documentclass[preprint,12pt]{elsarticle}

\usepackage{graphicx}
\usepackage{pdfpages}
\usepackage{subfigure}
\usepackage{float}
\usepackage{lineno}
\usepackage{amsmath}
\usepackage{multirow}

\journal{Nuclear Instruments and Methods in Physics Research A}

\biboptions{sort&compress}

\begin{document}
	
\begin{frontmatter}
			
	\title{Measurement of radon concentration in the output water of the 100~t/h ultrapure water system at the Jiangmen Underground Neutrino Observatory }
	
	\renewcommand{\thefootnote}{\fnsymbol{footnote}}

\author{
		C.B.Z.~Luo$^{a}$,
		Q.~Tang$^{a},\footnote{Corresponding author. E-mail address: tanquan528@sina.com (Q.~Tang).}$,
		C.~Guo$^{b,c,d}\footnote{Corresponding author. E-mail address: guocong@ihep.ac.cn (C.~Guo).}$,
		B.~Wang$^{a}$,
		J.C.~Liu$^{b,c,d}$,
		Y.P.~Zhang$^{b,c,d}$,
		L.D.~Lv$^{a}$,
		L.P.~Xiang$^{a}\footnote{Corresponding author. E-mail address: elephantxlp@163.com (L.P.~Xiang).}$,
		C.G.~Yang$^{b,c,d}$,
		B.~Xiao$^{e}$
	}

\address
{
	${^a}$ School of Nuclear Science and Technology, University of South China, Hengyang, 421001, China\\
	${^b}$ Experimental Physics Division, Institute of High Energy Physics, Chinese Academy of Sciences, Beijing, 100049, China \\
	${^c}$ School of Physics, University of Chinese Academy of Sciences, Beijing, 100049, China \\
	${^d}$ State Key Laboratory of Particle Detection and Electronics, Beijing, 100049, China \\
	${^e}$ School of Mechanics and Optoelectronic Physics, Anhui University of Science and Technology, Huainan, 232001, China. \\
}

	\begin{abstract}
	The Jiangmen Underground Neutrino Observatory (JUNO), a 20 kton multi-purpose low background liquid scintillator detector, was proposed primarily to determine the neutrino mass ordering. To mitigate radioactivity from surrounding rock and enable cosmic muon tagging, its central detector is immersed in a Water Cherenkov Detector (WCD) containing 40~ktons of ultrapure water instrumented with 2400 20-inch micro-channel plate photomultiplier tubes. Stringent radiopurity requirements mandate a radon concentration below 10 ~mBq/m$^3$ in the WCD. To achieve this, we developed a two-stage (ground and underground) ultrapure water system with 100~t/h production capacity, integrating a five-stage degassing membrane for radon removal. A novel microbubble technique was implemented to optimize the degassing membranes' radon removal efficiency. The synergistic combination of the microbubble technology and the multistage degassing membranes achieved a radon removal efficiency exceeding 99.9\%, reducing the system's output to 0.61 $\pm$ 0.50~mBq/m$^3$ in recirculation mode, surpassing design specifications and establishing world-leading performance standards. This paper details the ultrapure system architecture, quantifies the radon contributions of each device, and presents a comprehensive study on microbubble-augmented membrane degassing for low radon ultra-pure water production in a 100~t/h water system.

	\end{abstract}
	
			\end{frontmatter}


	\section{Introduction}
	
	The Jiangmen Underground Neutrino Observatory (JUNO)~\cite{1} is a state-of-the-art neutrino physics experiment located in South China. With 20 ktons of ultrapure liquid scintillator, JUNO aims to achieve groundbreaking measurements, including the determination of neutrino mass ordering and the precise measurement of three neutrino oscillation parameters ($\sin^{2}\theta_{12}$, $\Delta m^{2}\,_{21}$) with sub-percent precision by detecting reactor antineutrinos released from the Yangjiang and Taishan nuclear power plants. The JUNO detector comprises a Central Detector (CD), a Water Cherenkov Detector (WCD), and a top tracker detector.	To suppress the radioactive background from the surrounding rocks and to tag cosmic ray muons, the CD is submerged in the WCD, which contains 40 ktons of ultrapure water and 2400 Multi-Channel Plate PhotoMultipliers (MCP-PMTs)~\cite{2}. Monte-Carlo (MC) simulation results identify the dissolved radon ($^{222}$Rn) gas in water as a significant contributor to the CD's accidental background, requiring reduction to less than 10~mBq/m$^3$ for negligible impact~\cite{1}.  

An online UltraPure Water (UPW) production and circulation system with a production capability of 100~t/h is designed to supply UPW for the WCD continuously.  To control the radon concentration in water, this system integrates a microbubble generator and 2 sets of degassing membrane units, and is equipped with a highly sensitive online radon concentration in water measurement device. This paper is structured as follows: Sec.~\ref{sec:Ultrapure water system} describes the 100~t/h UPW system, Sec.~\ref{sec:Radon concentration in water measurement} briefly describes the online radon concentration in water measurement system and the improvements made to the system compared to previous work, Sec.~\ref{sec:Results} presents the variations of radon concentrations in water across the water system, the synergistic performance of microbubble-enhanced membrane degassing, and results of the radon concentration in produced water, and Sec.~\ref{sec:Summary} summarizes key findings.

\section{UPW system}\label{sec:Ultrapure water system}

\subsection{System overview}

	\begin{figure*}[!htb]
	\centering 
	\includegraphics
	[width=14cm]{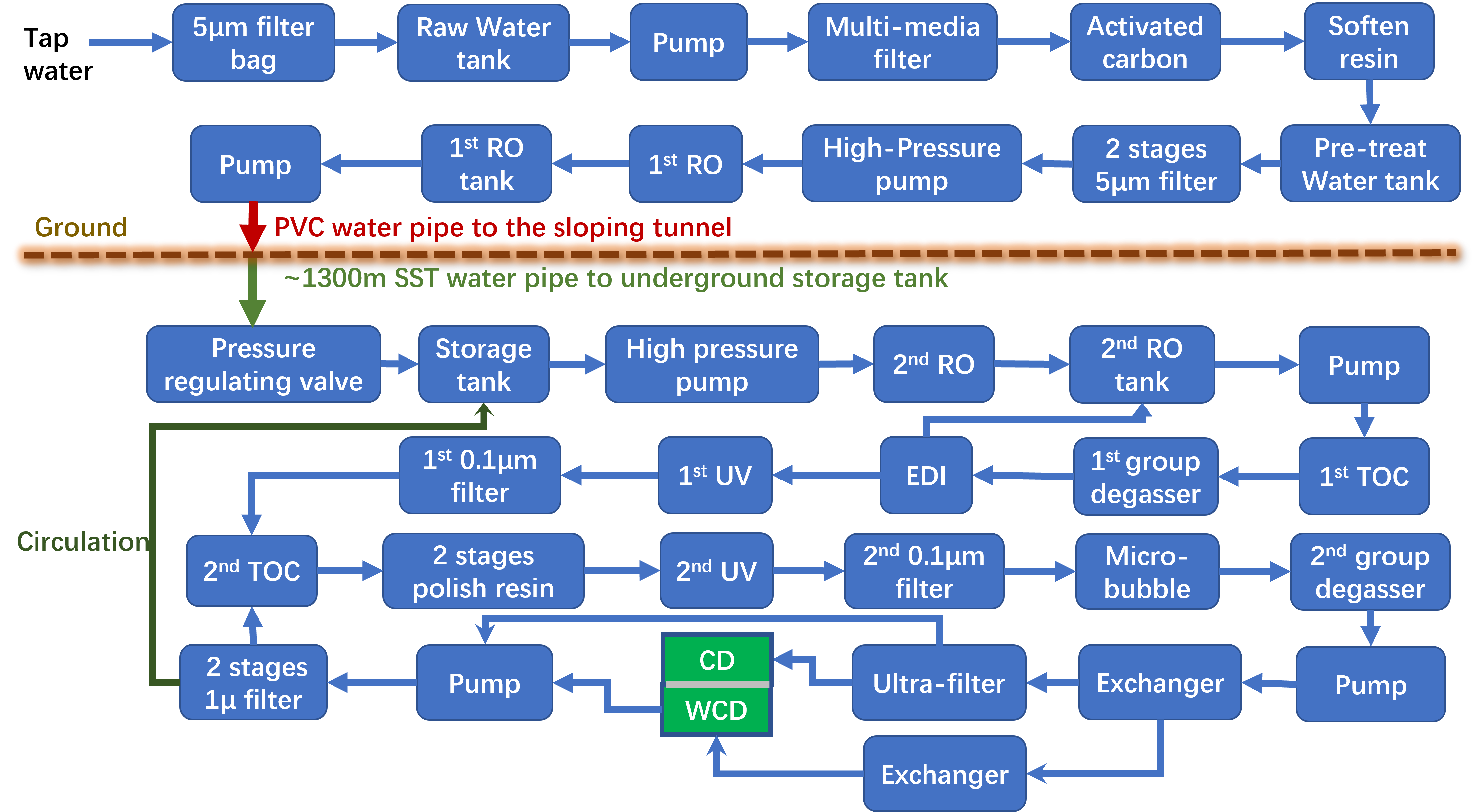}
	\caption{The flow chart of the JUNO 100~t/h UPW system. The system comprises aboveground and underground components, the produced water from the overground water system will be delivered to the underground water system as source water via a $\sim$1,300-meter Stainless STeel (SST) pipeline. Standard PVC pipes were used pre-EDI, while post-EDI, Sekisui Clean PVC pipes (Japan), specifically engineered for UPW conveyance, were implemented. }
	\label{fig:watersystem}
\end{figure*}

	The JUNO UPW system employs tap water as feedwater to produce UPW compliant with China's national electronic grade EW1 standard~\cite{3}. As illustrated in Fig.~\ref{fig:watersystem}, the system comprises aboveground and underground components, with the aboveground section featuring sequential treatment units: bag filters (5~$\mu$m pore size), multi-media filters (stratified beds of quartz sand, activated carbon, and anthracite), activated carbon filters (activated carbon made by coconut shell), softening resin columns (a special resin with high radium removal efficiency produced by a Chinese company), cartridge filters (5~$\mu$m pore size), and Reverse Osmosis (RO) membranes. The pretreatment sequence initiates with mechanical filtration through bag filters for gross particulate and sediment removal, followed by multi-media filtration for suspended solids reduction. Subsequent activated carbon adsorption eliminates residual chlorine, organic compounds, ammonia-nitrogen species, nitrites, trace heavy metals, and microbial contaminants. Ion-exchange softening achieves ions (such as Ca$^{2+}$ and Mg$^{2+}$) removal through cationic substitution.  Following pretreatment storage, the water passes through a cartridge filter (5~$\mu$m pore size) to intercept particulate matter and protect downstream high-pressure equipment. The RO membrane array provides primary desalination through semipermeable membrane technology, achieving near-complete ionic rejection efficiency~\cite{4}. The multi-media filters, activated carbon filters, and softening resin units are regenerable process components. Multi-media and activated carbon filters undergo periodic regeneration via hydraulic backflushing to restore filtration efficiency and adsorption capacity, respectively. The softening resin is regenerated through sodium chloride (NaCl) brine elution to reverse ion-exchange saturation, effectively displacing accumulated ions from the resin matrix and restoring its cationic exchange capacity. This regeneration protocol ensures operational longevity and sustainable performance across multiple service cycles.

The produced water from the overground water system will be delivered to the underground water system as source water via a 1,300-meter SST pipeline. The altitude difference between the two systems is $\sim$450 meters, so the water delivered to the underground first passes through a pressure-reducing valve before entering the water tank. The devices used in the underground water system are RO, Electrode De-Ionization (EDI), filters, Total Organic Carbon (TOC) removal, Ultra-Violet (UV), degassing membranes, microbubbles, heat exchangers, UltraFiltration (UF), etc. RO is the same as that used in the overground water system, and is used to remove ions from water further. EDI units achieve advanced ion removal through the synergistic operation of ion-selective exchange membranes and ion-exchange resins under a DC electric field. The process combines the selective permeability of anion/cation exchange membranes with the ion-exchange capacity of resin to enable directional ionic migration. This electrochemically driven hybrid mechanism ensures continuous ultrapure water production through permanent ionic impurity elimination without chemical regeneration requirements~\cite{5}. The pore size of the filter element used in the underground water system is 0.1~$\mu$m, and its main function is to remove the residual particles from water further. TOC and UV are both sterilized through ultraviolet irradiation and realize the oxidative decomposition of organic matter. The degassing membrane is mainly composed of hollow fibers, which are used with vacuum pumps to realize the gas removal from water with the help of a pressure difference~\cite{6}. The degassing membrane is the key equipment for removing radon from water~\cite{7,8,9}, and 2 groups of 5-stage degassing membranes are used to remove radon from water in this system. The main function of the microbubble generator is to load gas into the water, which can improve the degassing membranes' radon removal efficiency. Specific details of the degassing membranes and the microbubble generator will be introduced later. The heat exchanger is used to regulate the water temperature. The UF device is based on an advanced membrane separation technology, and its ultrafiltration membrane is microporous up to 10~nm or less, which can effectively remove particles, colloids, bacteria, etc. from water. 

Standard PVC pipes were used before the EDI module, while after that, Clean PVC pipes made by  Sekisui (Japan), specifically engineered for UPW conveyance, were implemented.

The UPW system performs dual functions: supplying UPW for both the WCD and the initial filling of the CD~\cite{1}. A critical design consideration necessitates separate treatment pathways due to differing operational requirements. The CD-bound water undergoes additional UF treatment to achieve a lower level of ionic contaminants, particularly radionuclides (U/Th/Ra)~\cite{10}, as it maintains direct interfacial contact with the liquid scintillator. The other path goes through an additional heat exchanger to lower the water temperature before entering the WCD, thus counteracting the heat generated by the PMT electronics immersed in it, and therefore maintaining a stable temperature throughout the CD. 

There are two operation modes for the UPW system, namely the filling mode and the recirculation mode. In the filling mode, the tap water passes through the various devices in turn and enters the CD or WCD. In the recirculation mode, the water in the WCD is pumped out, passed through a 1~$\mu$m filter, sent to the underground storage tank, and then returned to the WCD through the underground water system. Since this process generates a certain amount of wastewater, the lost water is replenished by the aboveground water system. With a 100~t/h recirculation flow rate, the wastewater volume of the underground system is $\sim$6~t/h.

\subsection{Radon removal devices}
The JUNO UPW system must not only satisfy conventional water purification standards but also suppress radon concentrations below 10~mBq/m$^3$. Radon removal from water is achieved through integrated degassing membrane modules and a microbubble generator.

\subsubsection{Degassing membrane}

\begin{figure}[htb]
	\centering
	\includegraphics[width=12cm]{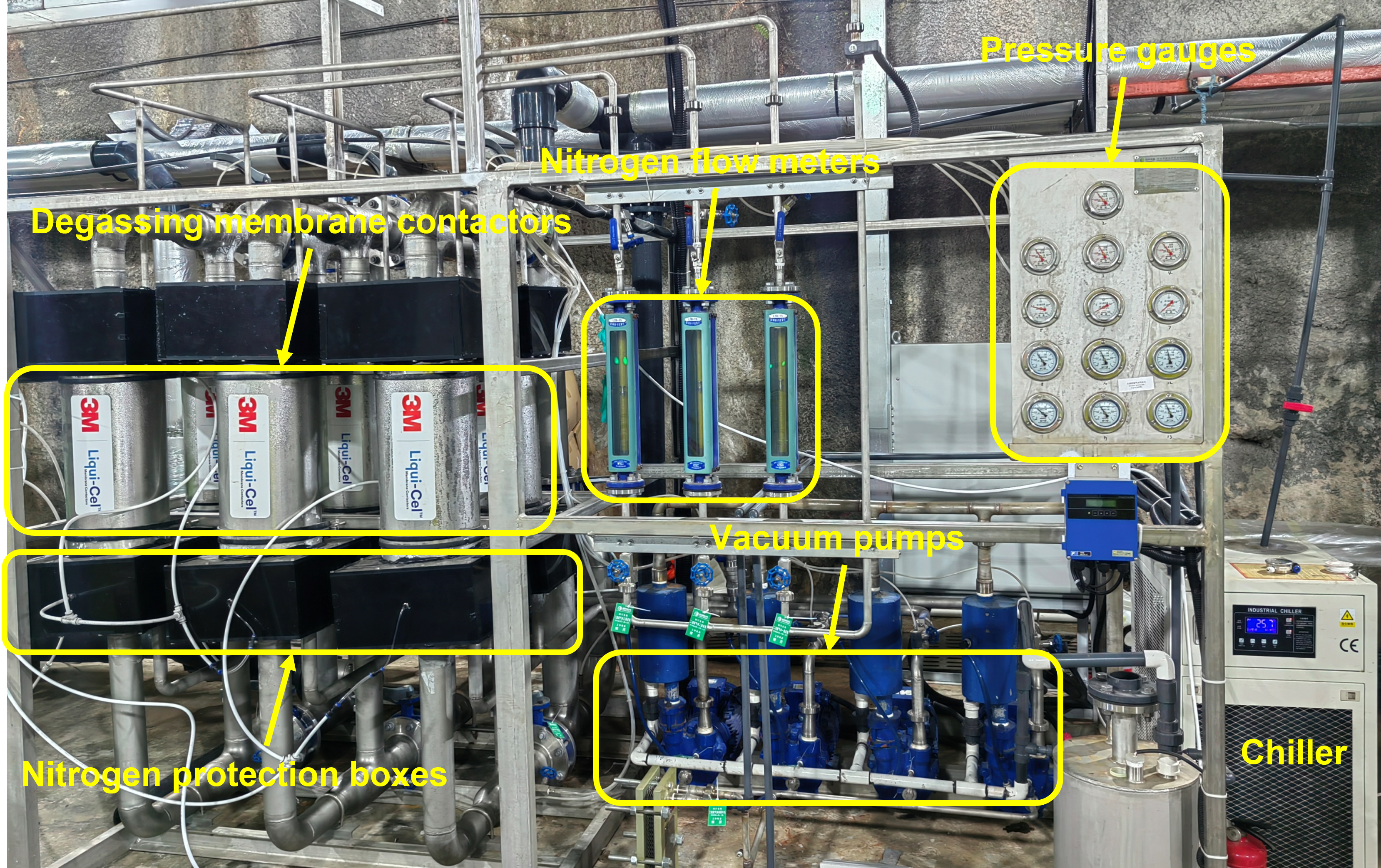}
	\caption{Picture of the second group of degassing membranes used in the JUNO UPW system.}
	\label{degasserpicture}
\end{figure}

The JUNO UPW system employs 2 groups of 3M$^{TM}$ Liqui-Cel$^{TM}$ degassing membranes. The first group operates in a two-stage configuration between the TOC unit and the EDI module, while the second group precedes the heat exchanger in a three-stage cascade. Fig.~\ref{degasserpicture} shows a photo of the second group of degassing membranes. The device comprises six degassing membranes contactors, four vacuum pumps, three nitrogen flow meters, one chiller, and several matching pressure gauges and valves. 

The degassing membrane contactors are the main structure of the unit, in which UPW flows, the vacuum pumps are used to provide a vacuum environment for the degassing membrane contactors, the flow meters ares used to control the flow rate of the purge gas (boil-off nitrogen used in this work), the chiller is used to dissipate the heat of the vacuum pumps, and the pressure gauges are used to monitor the pressure inside the degassing membrane contactors.

The degassing membrane contactor is a stainless steel chamber, sealed at the top and bottom with silicone rings and clips. The radon concentration in the underground environment of JUNO is $\sim$300~Bq/m$^3$~\cite{11}. Given the high penetrability of radon gas in silicone~\cite{12}, nitrogen protection boxes were used to avoid the penetration or leakage of radon into the degassing membrane contactors. Twelve nitrogen protection boxes with gas inlet and outlet ports and purged with boil-off nitrogen are wrapped around the ends of the contactors to prevent radon from entering them. The radon concentration in the nitrogen protection box is maintained at less than 1~Bq/m$^3$, which is measured by a RAD7 radon detector~\cite{35}.  

\begin{figure}[htb]
	\centering
	\includegraphics[width=10cm]{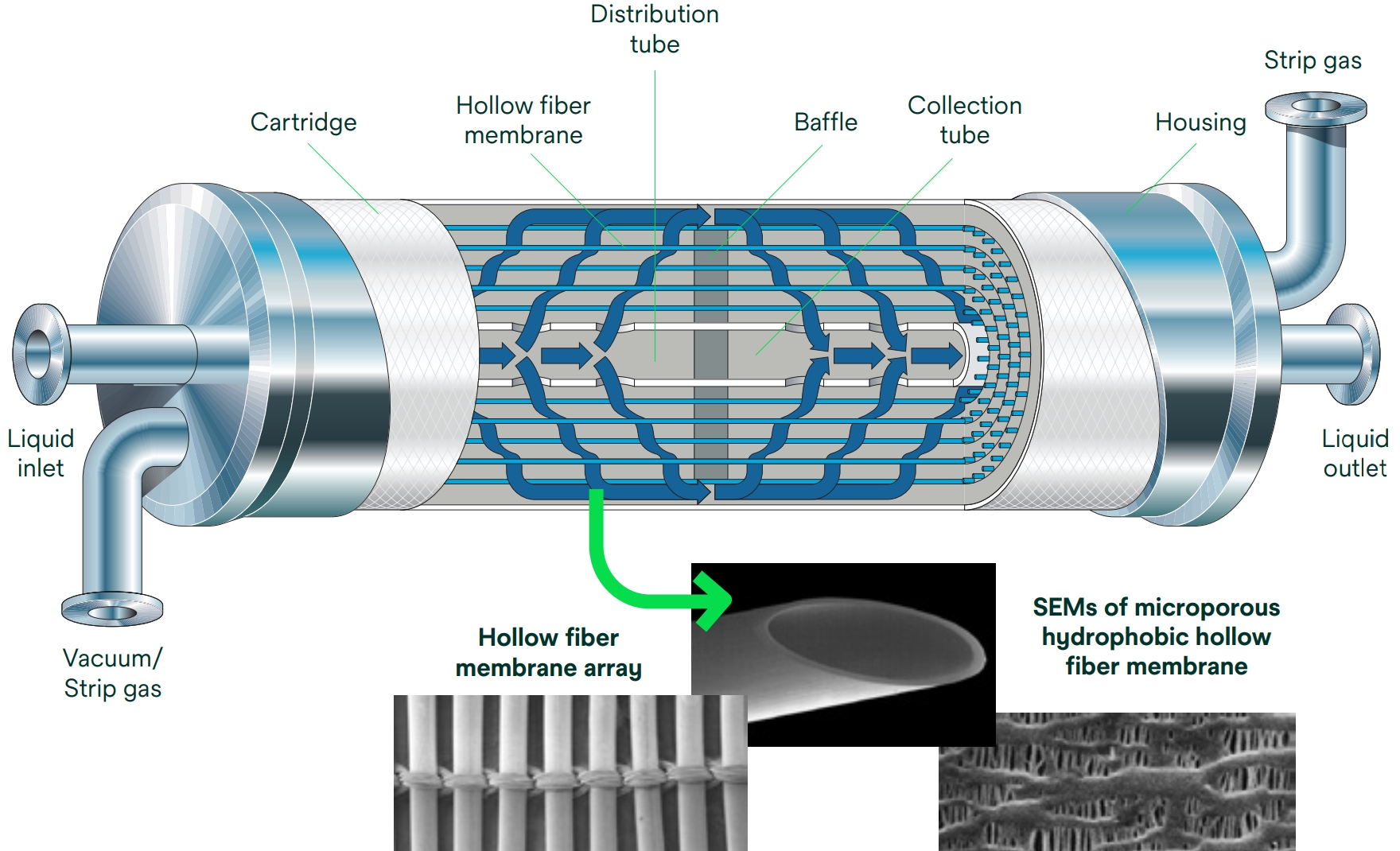}
	\caption{Schematic diagram of the degassing membrane.}
	\label{Schematic_diagram_degasser}
\end{figure}

Fig.~\ref{Schematic_diagram_degasser} shows the schematic diagram of the degassing membrane~\cite{13}. The degassing membrane module is based on microporous polypropylene hollow fiber membrane technology\cite{14,15,16}, which is comprised of an intricate arrangement of microporous polypropylene hollow fibers wound around a central tube. The core structure consists of hydrophobic hollow fibers uniformly dispersed in a helical configuration around the periphery of the center tube, which enhances the flow dynamics by using the special configuration of the microporous hollow fiber membrane and realizes the efficient stripping of dissolved gases through the extended membrane surface area. The fiber gap is precisely designed to balance the synergistic use of flow and membrane surface area, while the hydrophobic nature of the membrane effectively blocks liquid water penetration. When water passes through the inner side of the membrane, the vacuum pump establishes a negative pressure gradient at the outside of the fibers, driving dissolved gases (including radon, oxygen, and carbon dioxide) through the membrane holes and out of the system.

Liqui-Cel Membrane contactors utilize three modes of operation: vacuum, sweep gas, and hybrid synergy. In vacuum mode, a vacuum pump generates negative pressure in the fiber chamber, which causes dissolved gases in the liquid to migrate through the membrane pores to the other side of the fiber through the differential pressure and subsequently evacuate. Removal efficiency is positively correlated with water pressure and vacuum, consistent with Henry's Law, which states that an increase in differential pressure increases the gas transfer rate~\cite{14,17}. The gas sweep mode utilizes the principle that a high-purity carrier gas flows countercurrently through the outer side of the fiber tube to achieve a reduction in partial pressure by gas phase displacement. The hybrid synergistic mode integrates vacuum extraction with sweep gas purging. Experimental validation through our JUNO UPW system prototype demonstrated that this combined approach achieves optimal gas separation efficiency, contingent upon precise optimization of sweep gas flow rates~\cite{9,18}. Furthermore, our investigations revealed that the degassing membrane's radon removal efficiency depends not solely on the radon concentration but also on the total dissolved gas content in water\cite{8,18}. To increase the radon removal efficiency of the degassing membranes, we implemented a microbubble generator in the JUNO UPW system to enhance the gas content in water.

\subsubsection{Microbubble generator}
\begin{figure}[htb]
	\centering
	\includegraphics[width=6cm]{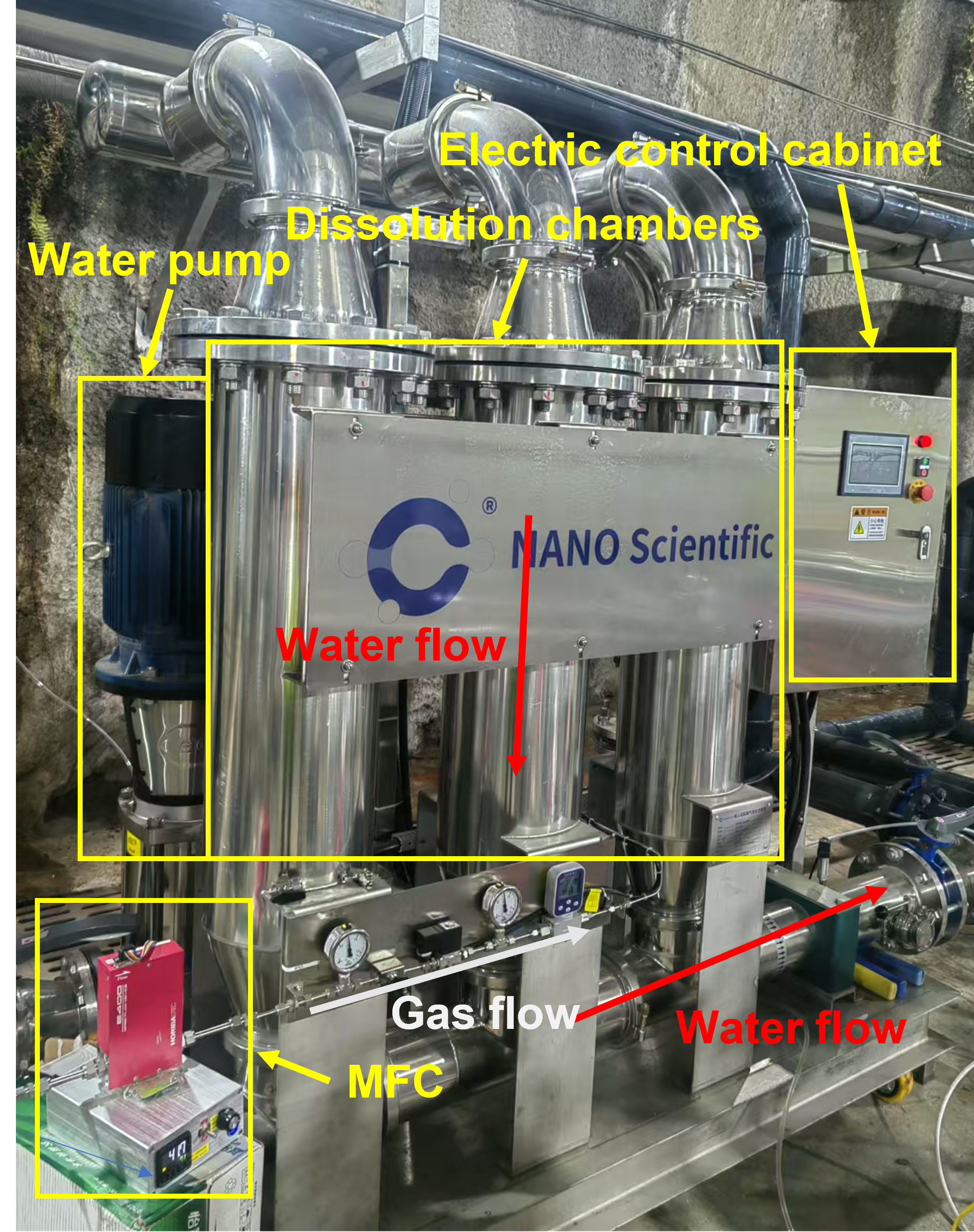}
	\caption{Picture of the microbubble generator used in the JUNO UPW system.}
	\label{mbpicture}
\end{figure}

Fig.~\ref{mbpicture} shows a picture of the microbubble generator used in the JUNO 100~t/h UPW system, which is made by Shanghai Xing Heng Technologies Co., Ltd. The microbubble generator unit consists of a water pump, three dissolution chambers, a Mass Flow Controller (MFC), an electric control cabinet, and supporting pipelines and valves. The basic principle of the device is dissolved gas release, i.e., the generation of microbubbles through the mechanism of over-pressurized dissolution and depressurized release\cite{19,20,21,22}. The system uses boil-off nitrogen with a pressure of 0.8~MPa as the gas source.

\begin{figure}[htb]
	\centering
	\includegraphics
	[width=8cm]{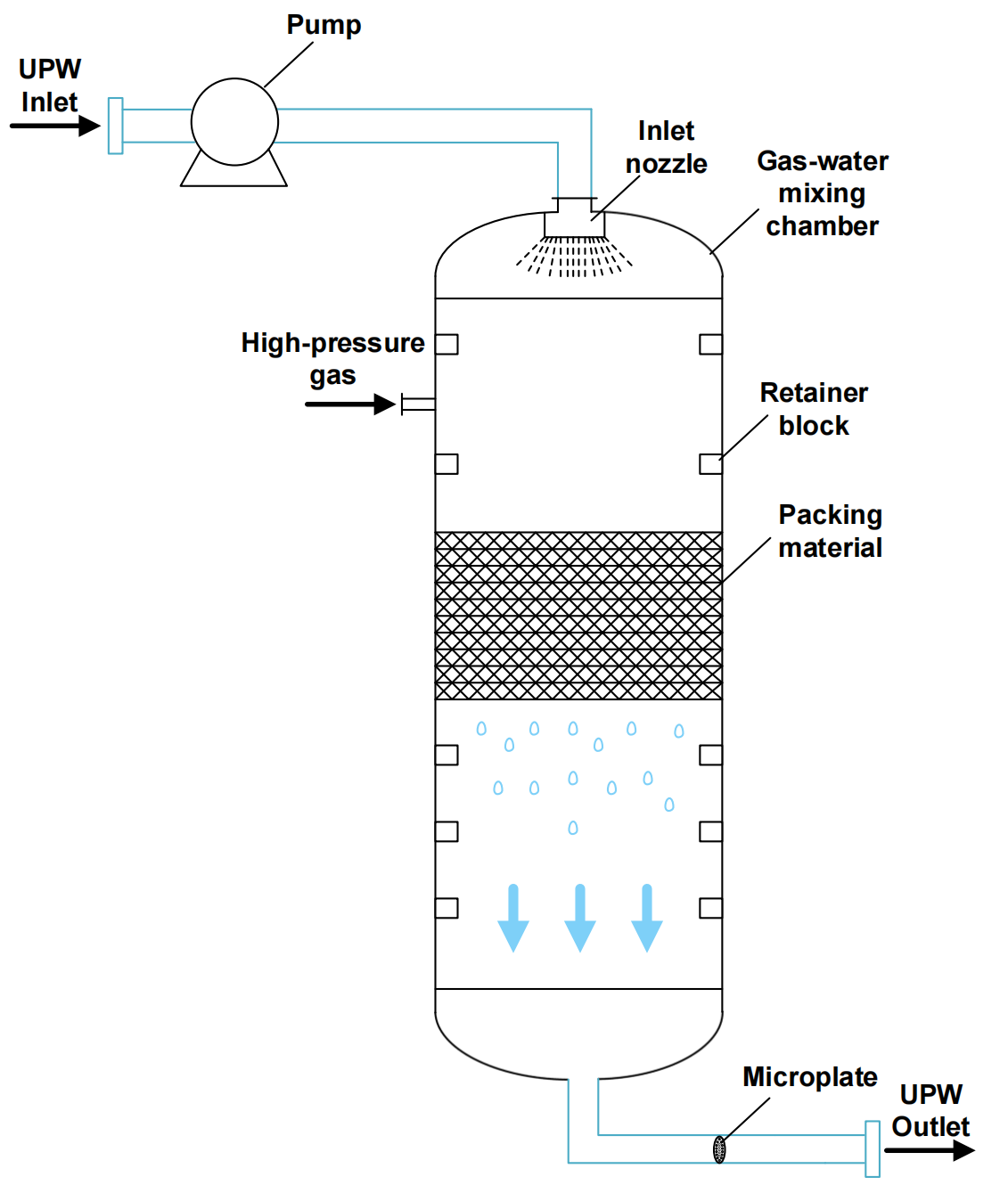}
	\caption{The schematic diagram of the microbubble generator.}
	\label{schematic_MB}
\end{figure}

\begin{figure}[htb]
	\centering
	\includegraphics
	[width=8cm]{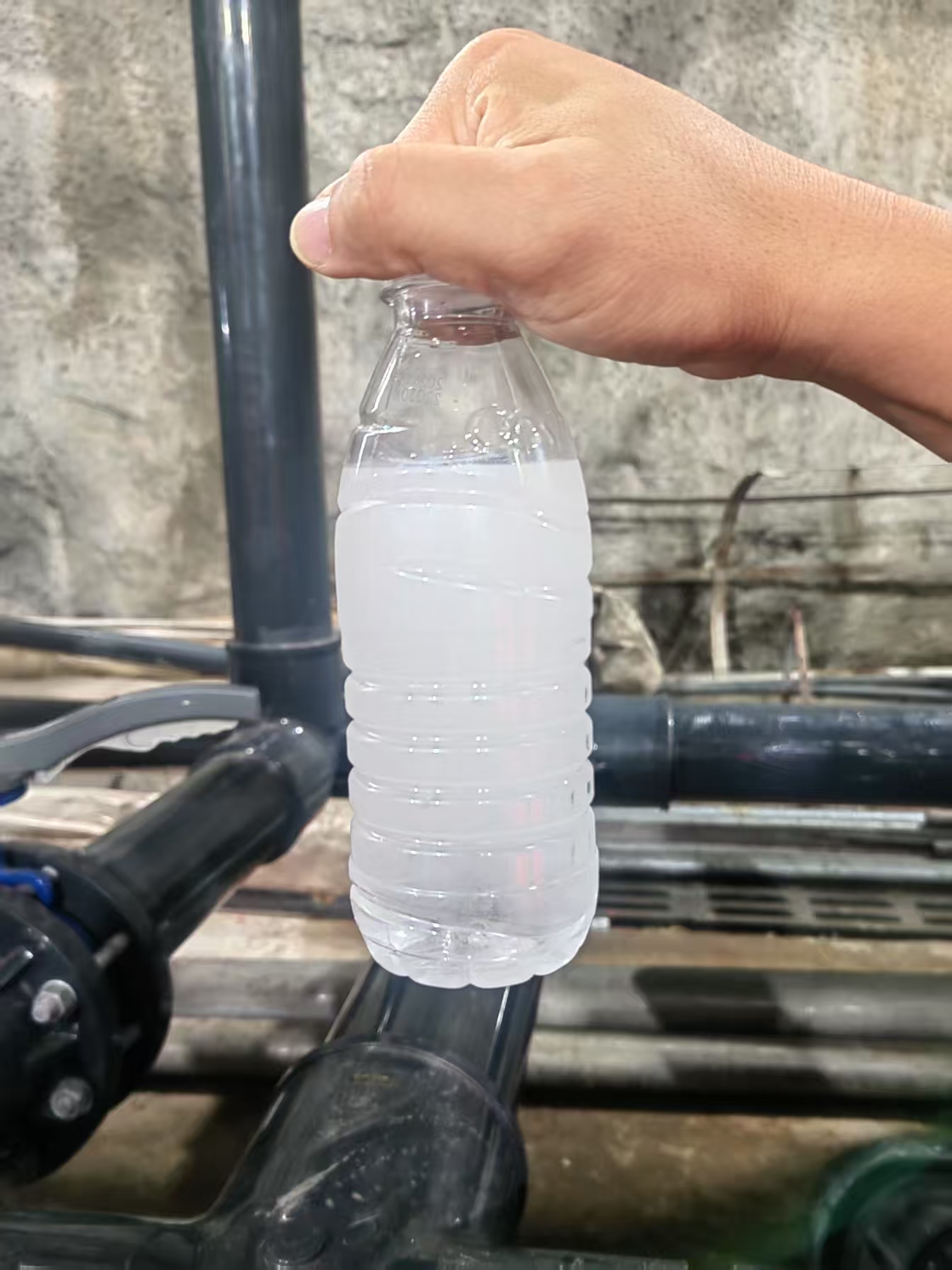}
	\caption{Picture of the micro-bubble-laden water sample.}
	\label{watersample}
\end{figure}

Fig.~\ref{schematic_MB} illustrates the experimental schematic of the microbubble generator. Ultrapure water is initially pressurized via a pump and subsequently injected into the dissolution chamber through a nozzle located at the top. Simultaneously, high-pressure nitrogen is introduced into the dissolution chamber through a side-mounted gas inlet. The chamber contains structured packing materials designed to maximize the gas-liquid interfacial area, thereby enhancing nitrogen dissolution efficiency. The nitrogen-loaded ultrapure water exits from the chamber’s bottom and undergoes rapid pressure reduction through a microplate, generating a dense population of microbubbles within the water. Fig.~\ref{watersample} displays the microbubble-laden water sample collected via pipeline sampling.

\section{Radon concentration in water measurement}\label{sec:Radon concentration in water measurement}
\subsection{Detection principle}
\begin{figure}[htb]
	\centering
	\includegraphics[width=10cm]{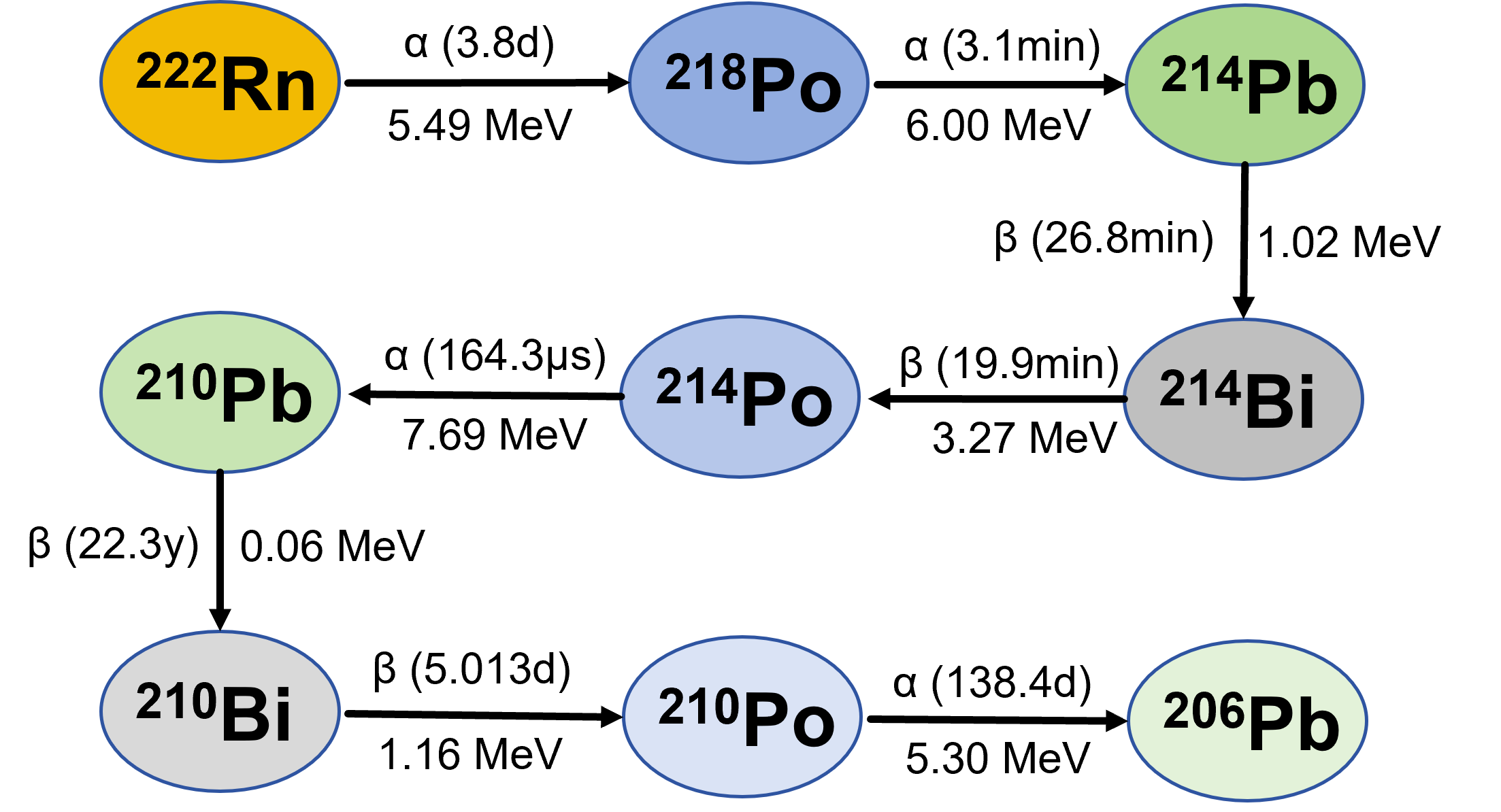}
	\caption{The relevant branch of $^{222}$Rn decay chain. The half-lives of the nuclide are in parentheses. The $\alpha$ decay energy or the maximum $\beta$ decay energy is indicated below the arrow.}
	\label{decaychain}
\end{figure}
$^{222}$Rn, a naturally occurring radioactive noble gas with a half-life of 3.82 days, constitutes the predominant background interference in radiation detectors due to its highest natural abundance among radon isotopes. The detection methodology exploits the characteristic $\alpha$s detected through its decay chain, with the correlated decay pathways systematically illustrated in Fig.~\ref{decaychain}. 

In this work, electrostatic collection measurements~\cite{23,24,25,26,27,28,29,30,31} are implemented to quantify $^{222}$Rn activity by spectroscopically detecting the monoenergetic $\alpha$s from its short-lived progeny $^{218}$Po and $^{214}$Po. Notably, this technique exclusively operates in gaseous-phase detection systems. Consequently, a critical prerequisite for accurately determining the $^{222}$Rn concentrations in water lies in the efficient radon gas transfer from water samples\cite{9}\cite{32}.

\subsection{System upgrade}

\begin{figure*}[htb]
	\centering
	\includegraphics
	[width=14cm]{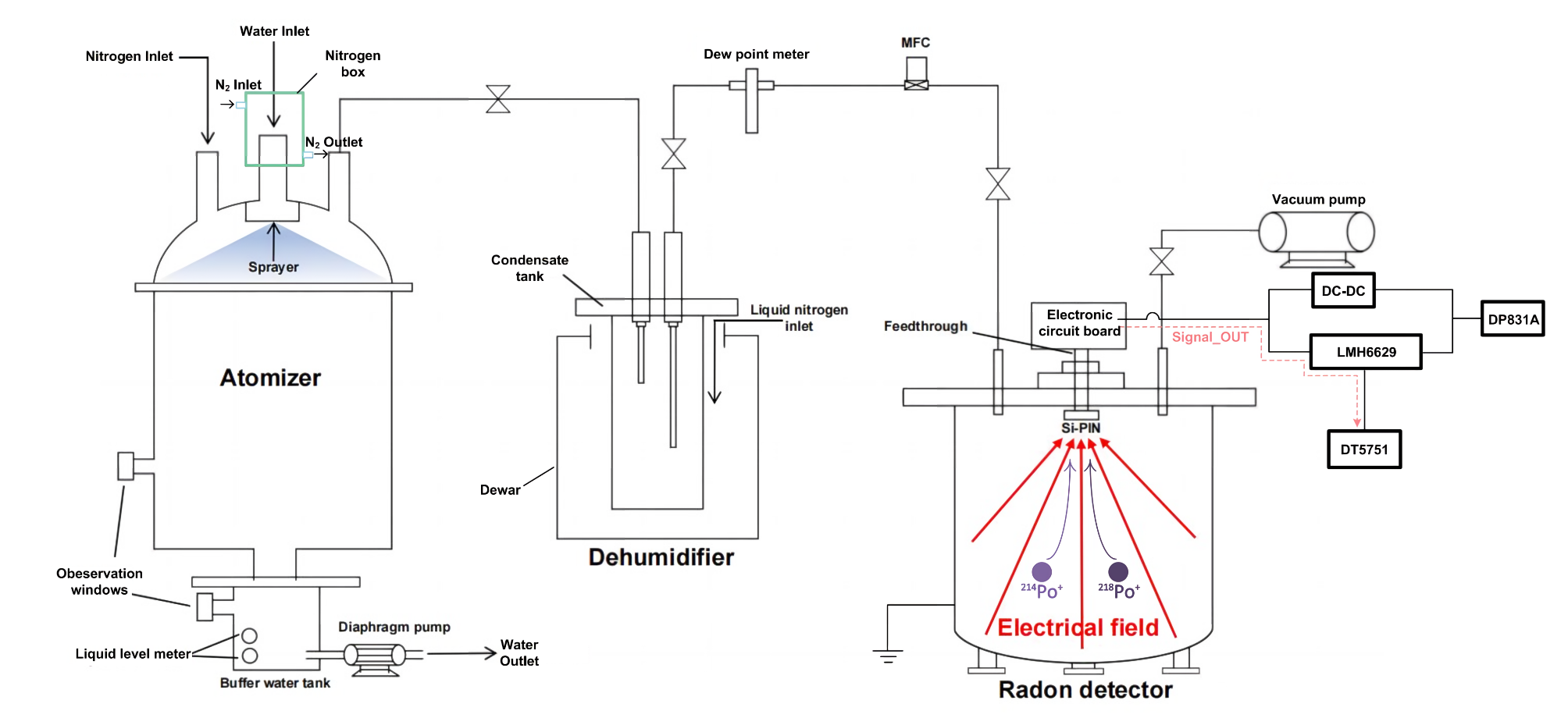}
	\caption{Schematic diagram of the radon concentration in water measurement system.}
	\label{measurementsystem}
\end{figure*}

Fig.~\ref{measurementsystem} shows a schematic diagram of the radon concentration in water measurement system, which consists of an atomizer, a dehumidifier, a radon detector, and the supporting pipes and valves. The system has been described in detail in Ref.~\cite{9}, we only give a brief description and report the system's improvements here.

\subsubsection{Background reduction}

During the radon concentration measurement at the JUNO site, PolyEthylene (PE) pipes and a quick-plug connector are used to connect the sample valve of the UPW system and the atomizer. A nitrogen protection box is used here to keep the connector in a low-radon environment, thereby avoiding potential air leakage from the quick-plug connector that could affect the measurement results. The nitrogen protection box here is similar to the one used at the degassing membrane. The measurement result shows that the radon concentration in the outgas of the box is below 1~Bq/m$^3$, which is measured by RAD 7.

Compared to the previous work~\cite{9}, the inner surface of this detector was repaired for previous defects (tiny holes on the inner surface), and the inner surface was electrolytically re-polished. The filled gas for background measurement has been changed to high-purity nitrogen from boil-off nitrogen, which has a lower radon concentration~\cite{24,25}. With these improvements, the $^{214}$Po background event rate of the detector was reduced to $\sim$1 Count Per Day (CPD).

In our previous configuration, gas transfer from the atomizer to the radon detector was facilitated using a diaphragm pump (N022AT, KNF) with G-threaded connections at both the inlet and outlet ports. However, operational testing revealed minor gas leakage at the G-thread interfaces. Consequently, in this work, we eliminated the diaphragm pump and implemented a vacuum pressure differential method for gas transfer.

Before gas transfer operations, the radon detector was pumped to a vacuum, and the atomizer was pressurized to 0.1~MPa with high-purity nitrogen. And subsequent valve activation enables gas transfer controlled by the flow meter. The reason for filling 0.1~MPa high-purity nitrogen is that the voltage between the SiPIN and the SST chamber of the radon detector is $\sim$2200~V, and the low gas pressure will cause spark discharge. Filling the measurement system with 0.1~MPa of high-purity nitrogen gas beforehand ensures that it is exactly at atmospheric pressure inside the detector after the gas transfer is completed. The reason for evacuating the detector is to provide the differential air pressure for the gas transfer.

In this system, metal-sealed with VCR or CF flange connections are used except for the atomizer water inlet, which uses a quick-plug connector, and the vacuum pump inlet, which uses a KF-25 vacuum flange connection. Despite the KF-25 flange utilizing a rubber O-ring seal in which the radon diffusion length is $\sim$2~mm~\cite{34}, no additional nitrogen protection box is implemented here, as a VCR valve is installed upstream of the extractor port.


\begin{figure}[htb]
	\centering
	\includegraphics
	[width=8cm]{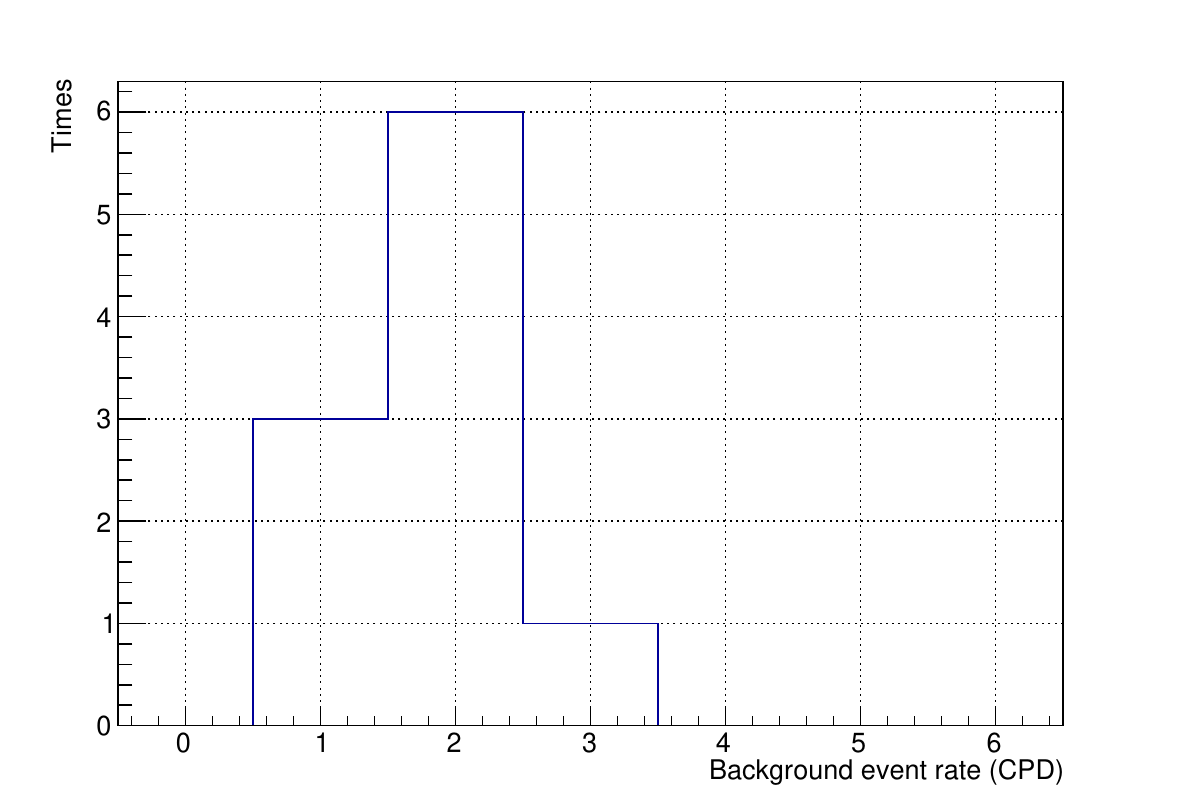}
	\caption{Distribution of the background event rate.}
	\label{background}
\end{figure}

In addition, it has been realized that the duration of each measurement is one day, so we do not choose to last the background measurement for many days at a time, but to do one-day background measurements many times. In the background measurement, high-purity nitrogen is first used to purge the system for 5 volumes, and then isolate the atomizer from other devices, fill the atomizer with 0.1~MPa high-purity nitrogen, vacuum the radon detector, and then transfer the gas from the atomizer to the detector at a flow rate of 1~L/min. The detector data-taking time for each measurement is 24 hours. The $^{214}$Po event rate recorded by the digitizer is the background result of a single measurement. In the present work, we performed 10 background measurements during the radon concentration in water measurement for the JUNO WCD system, the distribution of which is shown in Fig.~\ref{background}, and the average background event rate was 1.80 ± 1.34 CPD.

\subsubsection{Sensitivity estimation}

The sensitivity was estimated by measuring the system's intrinsic background and the detector's detection efficiency. At 90\% confidence level, the sensitivity can be estimated according to Eq.~\ref{sensitivity}~\cite{31}:
\begin{equation}\label{sensitivity}
L_{\text{c}} = \frac{1.64 \times \sigma_{\mathrm{BG}} \times R \times D}{ C_{\text{F}}}
\end{equation}
where L$_c$ is the sensitivity in the unit of mBq/m$^3$, $\sigma_{BG}$ is the statistical uncertainty of the system background in the unit of CPD, R is the ratio of radon concentration in water to gas, C$_F$ is the calibration factor of the detector in the unit of CPD/(mBq/m$^3$) which could reveal the detector's detection efficiency, D is the dilution factor which is 2 in this measurement. 

The relative humidity of the gas coming out of the atomizer is 100\%. As an electronegative gas, water vapor will combine with the positively charged $^{222}$Rn daughters, which will reduce the detector's collection efficiency. Based on our previous calibration results~\cite{9}, an operating temperature of -120~$^\circ$C was chosen in this work. At this temperature, the dehumidifier can reduce the absolute humidity of the measured gas to 7~mg/m$^3$ when the gas flow rate is 1~L/min. With these parameters, the detector's calibration factor is 2.59 $\pm$ 0.19 CPD/(mBq/m$^3$), corresponding to a $^{214}$Po detecting efficiency of $\sim$34\%.

Calculated according to Eq.~\ref{sensitivity} with the specific parameters, the system's one-day measurement sensitivity is 0.42~mBq/m$^3$.

\subsection{Radon concentration in water determination}

Following the gas transfer process, the radon detector is sealed, and a 24-hour measurement cycle is initiated. The DT 5751 digitizer records the signal waveforms, which are subsequently processed via offline analysis software to extract the $^{214}$Po event rate. The radon concentration in water is then calculated using Eq.~\ref{eq3}:
\begin{equation}
\label{eq3}
	C_{\text{Rn}} = \frac{(R_{\text{Po-214}}-R_{\text{b}}) \times D \times R }{ C_{\text{F}}}
\end{equation}
where C$_{Rn}$ is the radon concentration in water in the unit of mBq/m$^3$, R$_{Po-214}$ is the $^{214}$Po event rate for the sample measurement in the unit of CPD, R$_b$ is the system background event rate in the unit of CPD, D is the dillution factor which is 2 in this measurement, R is the radon concentration ratio in water to vapor which is 0.25 in this measurement, C$_F$ is the calibration factor.

\section{Results}\label{sec:Results}

\subsection{Radon contributions of the UPW system}
The JUNO UPW system uses tap water as the source water to produce UPW with a radon concentration less than 10~mBq/m$^3$. The system implements an innovative synergistic radon removal process combining degassing membranes with a microbubble generator. A comprehensive measurement of the radon contribution of the UPW system components has been conducted to quantitatively study the radon removal efficiency of the degassing membrane and the microbubble generator, as well as to optimize and improve the water system's radon removal efficiency. Tab.~\ref{tab.UPWRn} shows the radon concentration measurement results at different locations of the UPW system, which works in the filling mode. The overground samples were measured using a RAD7 radon detector, while the underground measurements were conducted with our custom-built system. The errors of  S7 and S8 are statistical only.

\begin{table*}[htb]
	\caption{Measurement results of radon concentration (C$_{^{222}Rn}$) at different places of the UPW system. The UPW system works in the filling mode. S1-S6 were measured using a RAD7 detector. S7 and S8 were measured by our custom-built system. The errors for S7 and S8 include both statistical and systematic errors.}
	\centering
	\renewcommand{\arraystretch}{1.3}
	\normalsize
	\begin{tabular} {@{\extracolsep{\fill} } ccc}
		\hline
		\multicolumn{1}{c}{Sample}&{C$_{^{222}Rn}$}&{ Location} \\
			& (Bq/m$^{3}$) & \\
		\hline
		S1  & 6.1 $\pm$ 3.1 & Tap water\\
		S2  & 6.3 $\pm$ 2.5 & After 5$\mu$m filter bag \\
		S3  & 9.7 $\pm$ 2.0  & After Multi-media filter \\
		S4  & 11.4 $\pm$ 2.5 & After Activated carbon filter\\
		S5  & 16.9 $\pm$ 2.5  & After the Soften resin \\
		S6  & 10.0 $\pm$ 5.1  & After 1$^{st}$ stage RO \\
		S7  & 11.5 $\pm$ 0.9  & After the underground water tank \\
		S8  & 10.6 $\pm$ 0.8  & After the 2$^{nd}$ stage RO\\
		\hline
	\end{tabular}
	\label{tab.UPWRn}
	\end{table*}

The radon concentration in the tap water at the JUNO site is $\sim$6~Bq/m$^3$, however, subsequent treatment through multi-media filtration, activated carbon filtration, and ion-exchange softening resin induced a measurable increase, attributed to radon emanation from filler material.

The marked reduction from S5 to S6 stems from two mechanisms: 1) the radon degassing mechanism during temporary water storage, top-feed water injection generates turbulent gas-liquid interfaces between incoming and resident water, driving dissolved radon partition into the headspace air; 2) particle adsorption effect, RO feed water contains particles, which can adsorb a certain amount of radon, and RO removes some of the radon gas while removing the particles.

From the ground to the underground, the ultrapure water went through 1300~m of SST pipeline, which was cleaned in detail, so basically there was no contamination of the ultrapure water. The increase of S7 relative to S6 was still mainly caused by the turbulence generated by the new water entering the tank. Unlike what was mentioned before, the radon concentration in the upper gas of the underground tank was $\sim$100~Bq/m$^3$, which was higher than the equilibrium concentration ($\sim$40~Bq/m$^3$), so the radon entered the water from the gas. The reason for the slight decrease in S8 was related to the removal of particles by the RO.

Considering all the above factors, the radon concentration in the water entering the first group of degassing membranes was $\sim$10~Bq/m$^3$.

\subsection{Optimization of the degassing membranes}

Before microbubble generator implementation, operational optimization of the degassing membrane unit is essential. The degassing membranes worked in hybrid synergy mode. There were two optimizable quantities for the degassing membrane: one was the water pressure, and the other was the gas purge flow rate. 

Water pressure optimization required balancing the radon removal efficiency with the UPW production capability. Operating below 0.30~MPa resulted in insufficient UPW production, while exceeding 0.55 MPa risked compromising piping integrity. Given fixed-frequency pumps upstream of the first group of degassing membranes, pressure regulation was implemented at the secondary group by adjusting the frequency of the relevant pump, and the pressure optimization range was 0.30~MPa - 0.55~MPa.

Boil-off nitrogen was used as the purge gas, and its flow rate needed to be balanced between purge efficiency and vacuum maintenance. Insufficient flow affected the radon purge, while too high a flow rate reduced the vacuum on the outside of the fibers and therefore reduced the radon removal efficiency. The system allowed continuous flow rate adjustment employing a flow meter in the range of 0-10~m$^3$/h.

After each stage of degassing membranes, the radon concentrations in water were first measured under the condition of 0.40~MPa water pressure and 5~m$^3$/h nitrogen purge flow rate, and the results are shown in Tab.~\ref{tab.degasserRn}. In the final produced water, the radon concentration was 510 $\pm$ 30~mBq/m$^3$, which was far from the 10~mBq/m$^3$ experimental requirement.

\begin{table*}[htb]
	\caption{Radon concentrations in water after each stage of degassing membranes. The radon removal efficiency is calculated according to the $^{222}$Rn concentrations before and after each device. During this measurement, the water pressure of the first group of degassing membrane is 0.55~MPa, the water pressure of the second group of degassing membrane is 0.40~MPa, and the nitrogen purge flow rate is 5~m$^3$/h. }
	\centering
	\renewcommand{\arraystretch}{1.3}
	\normalsize
	\begin{tabular} {@{\extracolsep{\fill} } cccc}
		\hline
		\multicolumn{1}{c}{Sample}&{C$_{^{222}Rn}$}& {Radon removal efficiency}&{ Location} \\
			& (Bq/m$^{3}$) & (\%) & \\
		\hline
		S1  & 1.91 $\pm$ 0.15 & $\sim$82 & After 1$^{st}$ stage \\
		S2  & 1.24 $\pm$ 0.10 & $\sim$35 & After 2$^{nd}$ stage \\
		S3  & 0.88 $\pm$ 0.08 & $\sim$29 & After 3$^{rd}$ stage \\
		S4  & 0.68 $\pm$ 0.06 & $\sim$23 & After 4$^{th}$ stage\\
		S5  & 0.51 $\pm$ 0.05 & $\sim$25 & After 5$^{th}$ stage \\
		\hline
		\end{tabular}
		\label{tab.degasserRn}
		\end{table*}

\subsubsection{Water pressure optimization}
The primary degassing membrane group operated with an inlet pressure of 0.55~MPa, maintained by fixed-frequency pumps in the upstream system. This module exhibited an outlet pressure of 0.45~MPa, demonstrating a characteristic pressure drop of $\sim$0.05~MPa for each stage of the degassing membrane. The secondary degassing membrane system was pressure-regulated by adjusting the frequency of variable frequency pumps located upstream and downstream, and the inlet pressure was used as the main control parameter for system optimization. 

\begin{figure}[htb]
	\centering
	\includegraphics
	[width=8cm]{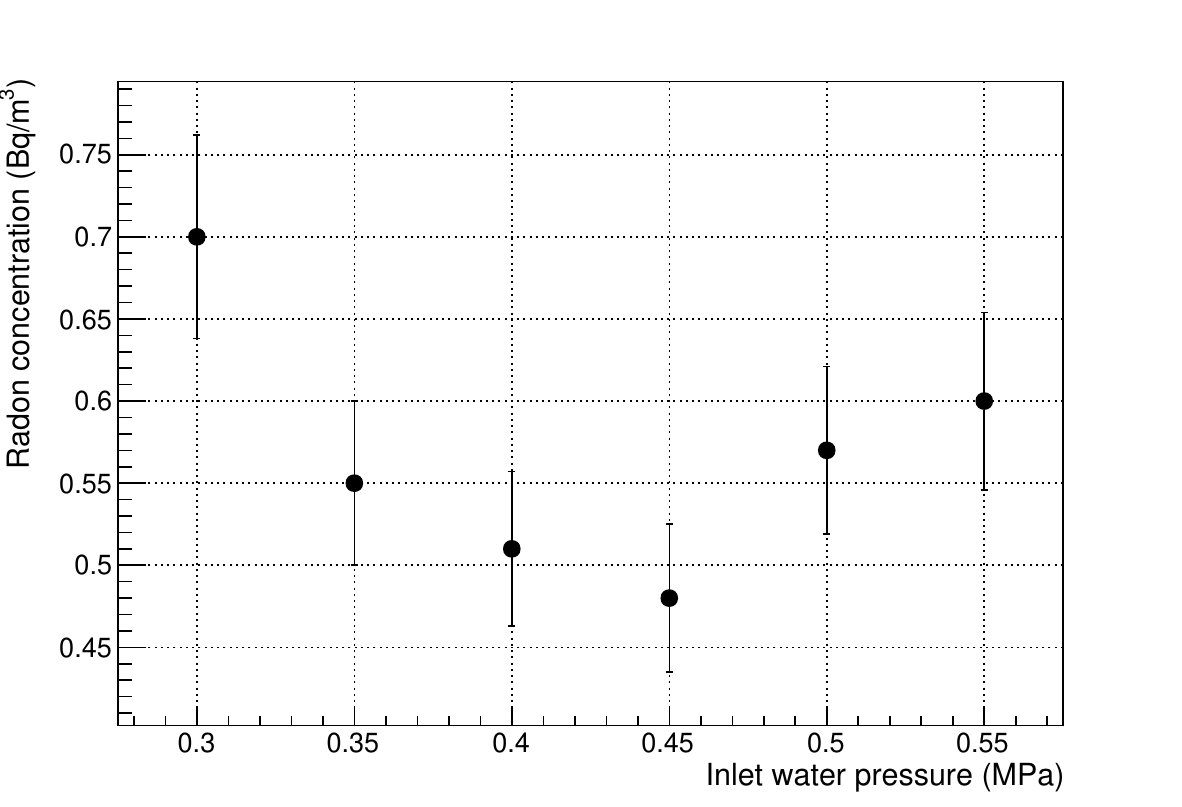}
	\caption{Radon concentration variation after the fifth-stage degassing membrane versus the secondary degassing module inlet pressure. The errors in the Y-axis include both statistical and systematic errors.}
	\label{pressureRn}
\end{figure}

Fig.~\ref{pressureRn} demonstrates the correlation between the radon concentration after the fifth-stage degassing membrane and the inlet water pressure of the secondary degassing module. Due to the system pressure-flow relationship, radon concentration did not consistently decrease as the inlet pressure increases: an inlet pressure of 0.35~MPa corresponded to a production capacity of $\sim$80~t/h, while 0.55~MPa was $\sim$100~t/h. The increase in flow rate shortened the gas-liquid contact time in the membrane, resulting in incomplete degassing. Experimental data showed that optimum performance was achieved at an inlet pressure of 0.45~MPa.

\subsubsection{Purge gas flow rate optimization}
\begin{figure}[htb]
	\centering
	\includegraphics
	[width=8cm]{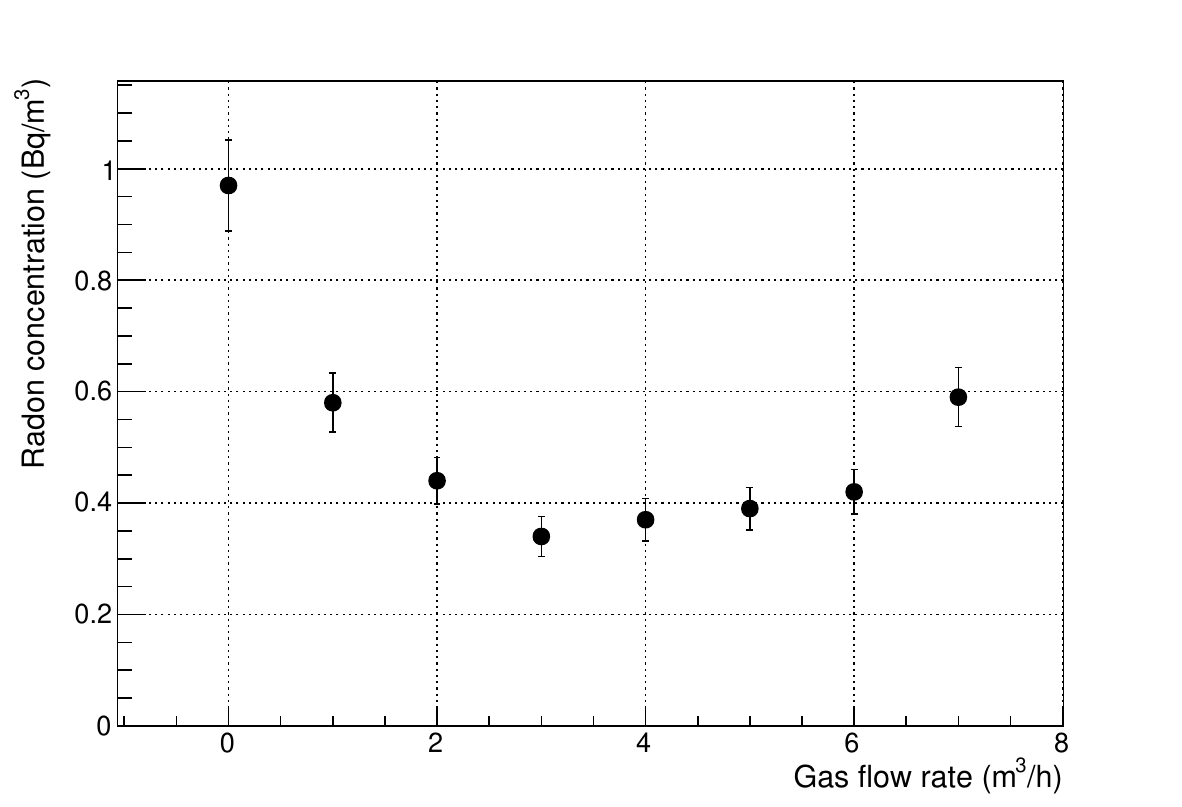}
	\caption{Radon concentration variation after the fifth-stage degassing membrane versus the purging nitrogen flow rate. The errors in the Y-axis include both statistical and systematic errors.}
	\label{gasRn}
\end{figure}

Flow meters preceding each degassing membrane stage maintained consistent nitrogen purge flow rates across all units, adjustable within a 0-10~m$^3$/h range. Similar to the water pressure optimization, the radon concentration after the fifth-stage degassing membrane served as the parameter for optimization. The results are shown in Fig.~\ref{gasRn}. 

Zero nitrogen purge flow rate indicated vacuum-mode operation of the degassing membrane, during which the system demonstrated suboptimal radon removal efficiency. However, higher nitrogen flow rates did not proportionally enhance performance due to their vacuum-compromising effect inside the contactor. As is evidenced by Fig.~\ref{gasRn}, optimal radon separation was achieved at 3~m$^3$/h nitrogen flow rate.

Without the use of a microbubble generator, the radon concentration in this UPW system can be reduced to a minimum of 340 $\pm$ 36~mBq/m$^3$. At this time, the inlet water pressure of the first degassing membrane unit was 0.55~MPa, and that of the second degassing membrane unit was 0.45~MPa. The degassing membranes worked in the hybrid synthetic mode with a nitrogen purging flow rate of 3~m$^3$/h.

\subsection{Combined commissioning of the degassing membranes and the microbubble generator}
\begin{figure}[htb]
	\centering
	\includegraphics
	[width=8cm]{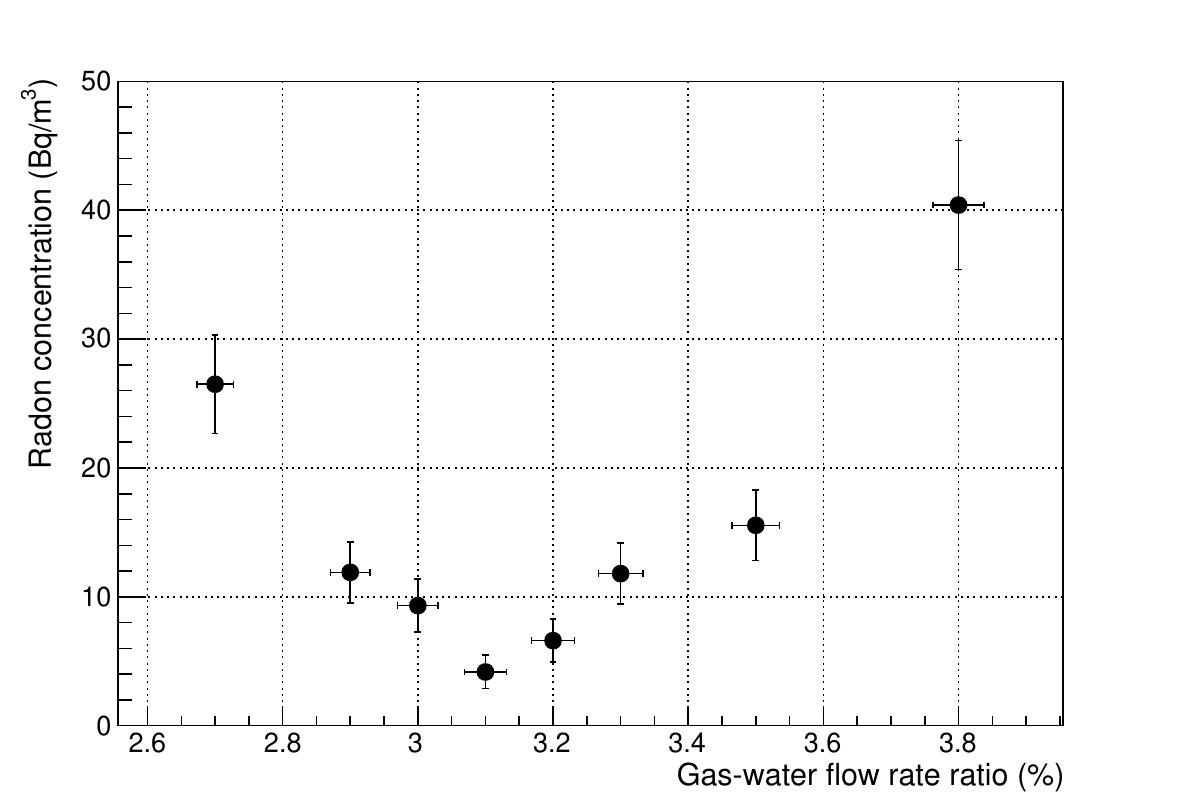}
	\caption{Radon concentration variation after the fifth-stage degassing membrane versus the gas-water flow rate ratio of the microbubble generator. The errors in the Y-axis are statistical only; X-axis error includes system flow uncertainty ($\pm 2\%$).}
	\label{gasflow}
\end{figure}

Microbubble generation was integrated into the system when the degassing membranes were working optimally. The generator's sole operational parameter was the gas injection flow rate with a maximum valve of 12~m$^3$/h, which corresponded to a gas-water flow rate ratio of 12\%. System validation maintained the radon concentration after the fifth-stage membrane as the critical performance metric, with detailed data presented in Fig.~\ref{gasflow}. 

According to the test results, the radon removal capability of the system varied greatly with the gas-water flow rate ratio of the microbubble generator, and the optimal value in the JUNO UPW system was 3.1\%, which was equivalent to an inlet gas flow rate of 3.1~m$^3$/h.

\subsection{Radon removal limit}
The JUNO UPW system operated in two distinct modes: feed mode and recirculation mode. In the feed mode, municipal water was processed through overground and underground treatment trains for direct detector filling. The works described in the previous sections were carried out in this mode. Under optimized parameters, the system achieves single-pass radon reduction from 10.6 $\pm$ 0.2~Bq/m$^3$ to 4.2 $\pm$ 1.3~mBq/m$^3$, corresponding to a radon removal efficiency better than 99.9\%. 

The JUNO experiment completed the initial detector UPW filling on February 2, 2025, with the UPW system transitioning to recirculation mode since then. At the beginning of the cycle, the radon concentration in the WCD output water was measured to be 161 $\pm$ 14~mBq/m$^3$. When reprocessed as feedwater, the system achieved a reduced radon concentration of 0.61 $\pm$ 0.50~mBq/m$^3$, validating its capability for sustained generation of sub-mBq/m$^3$ ultrapure water in the recirculation mode.

\section{Summary}\label{sec:Summary}
JUNO is a multi-purpose neutrino experiment primarily designed to determine the neutrino mass ordering. To lower the accidental background in the center detector, the $^{222}$Rn concentration in the water of the surrounding WCD should be controlled to less the 10~mBq/m$^3$. 

The JUNO UPW system, operating at 100~t/h capacity, has implemented an innovative five-stage membrane degassing cascade synergized with a microbubble generator, achieving unprecedented radon suppression.  Operating at optimized parameters, 0.55~MPa water pressure for the primary degassing membrane unit and 0.45~MPa water pressure for the secondary degassing membrane unit, 3~m$^3$/h nitrogen purge flow rate for the degassing membranes, and 3.1\% gas-water flow rate ratio for the microbubble generator, the system reduced radon from 10.6 $\pm$ 0.2 Bq/m$^3$ to 4.2 $\pm$ 1.3~mBq/m$^3$ ($>$99.9\% radon removal efficiency) in single-pass mode. Transitioning to recirculation post-filling, it further demonstrated sub-mBq/m$^3$ low radon UPW production performance, lowering WCD UPW from 161 $\pm$ 14~mBq/m$^3$ to 0.61 $\pm$ 0.50~mBq/m$^3$, exceeding JUNO's 10~mBq/m$^3$ requirement.

This technological breakthrough establishes a novel paradigm for multi-kiloton-scale experiments requiring sub-mBq/m$^3$ low radon UPW, thereby securing the essential low-background conditions for realizing JUNO's physics discovery potential.


\section*{Acknowledgments}
This work is supported by the Strategic Priority Research Program of the Chinese Academy of Sciences (Grant No. XDA10011200) and the Youth Innovation Promotion Association of the Chinese Academy of Sciences (Grant No. 2023015).

\section*{Data availability}
Data will be made available on request.

\end{document}